# A self-contained algorithm for determination of solid-liquid equilibria in an alloy system

L. Yang [a,b], Y. Sun [a], Z. Ye [a], F. Zhang [a,*], M. I. Mendelev [a], C. Z. Wang [a,b], and K. M. Ho [a,b,c]

[a] *Ames Laboratory, US Department of Energy, Ames, Iowa 50011, USA*

[b] *Department of Physics, Iowa State University, Ames, Iowa 50011, USA*

[c] *Hefei National Laboratory for Physical Sciences at the Microscale and Department of Physics, University of Science and Technology of China, Hefei, Anhui 230026, China*

**Abstract:** We describe a self-contained procedure to evaluate the free energy of liquid and solid phases of an alloy system. The free energy of a single-element solid phase is calculated with thermodynamic integration using the Einstein crystal as the reference system. Then, free energy difference between the solid and liquid phases is calculated by Gibbs-Duhem integration. The central part of our method is the construction of a reversible alchemical path connecting a pure liquid and a liquid alloy to calculate the mixing enthalpy and entropy. We have applied the method to calculate the free energy of solid and liquid phases in the Al-Sm system. The driving force for fcc-Al nucleation in Al-Sm liquid and the melting curve for fcc-Al and $Al_3Sm$ are also calculated.

*Corresponding author at: Ames Laboratory, Ames, IA, 50011, USA. E-mail address: fzhang@ameslab.gov (F. Zhang)

## 1 Introduction

Reliable free energy for both solid and liquid phases is fundamental to achieving a microscopic understanding of freezing and melting phenomena, which remains a significant challenge in condensed matter physics and materials science [1]. The origin of the difficulty in free energy calculations is that it cannot be expressed as a simple average of a physical quantity over the phase space, and thus cannot be evaluated in a single simulation with a standard sampling technique, such as Monte Carlo (MC) or Molecular Dynamics (MD) [2]. On the other hand, the derivatives of the free energy are usually measurable quantities in direct MC or MD simulations. Based on this, the thermodynamic integration (TI) method outlines a practical way of computing the free energy by evaluating its derivative along a reversible path connecting the realistic system and a reference system. The integration of the derivative along this path gives the free energy difference between the two systems [2,3]. In principle, one can obtain the absolute free energy of solid and liquid phases by referencing to a harmonic crystal and the ideal gas, respectively, whose free energy can be analytically derived. While the harmonic crystal, such as the Einstein crystal [4,5], can provide a reliable reference for solid phases in most cases, the usage of the ideal gas as the reference system for liquids is more problematic in real applications, since rather complex integration paths have to be created in the supercritical regime to avoid vapor-liquid phase transitions. In addition, many phenomena of interest such as crystal nucleation and growth occur when the liquid becomes supercooled, when it behaves so differently from the ideal gas that one can hardly obtain the required accuracy by using the ideal gas as the starting point. Furthermore, it is generally not a good idea to treat the liquid and solid phases in separate frameworks when it is the free energy *difference* that controls phase stability [6]. Here, we construct an "alchemical" path to transform a pure liquid to a liquid alloy, and apply TI to evaluate the mixing enthalpy and entropy during the process. Similar methods were frequently used previously to analyze affinity change upon substitution of certain atoms or functional groups in chemical or biochemical systems[7–9]. This strategy, together with a reliable method of determining solid/liquid free energy difference in single-element systems (discussed below), forms a self-contained way of establishing phase equilibria in alloys.

For completeness, we also discuss the free energy calculation of single-element solid and liquid phases. The general strategy is as follows: first, we calculate the absolute free energy of the solid phase directly using an Einstein-crystal reference; next, we determine the free energy difference between the solid and liquid phases at a specific state point; and finally, we use Gibbs-Duhem integration [10] to extend to other state points. While the free energy difference at an arbitrary state point can be calculated by methods such as pseudosupercritical path integration [11,12], in this paper, we choose a special state point: the melting point, at which the free energy difference is zero. The accurate melting point is determined by monitoring the migration of a solid-liquid interface.

We choose the Al-Sm system for the current study, which is a typical member of Al-RE systems (RE: rare earth). At ~10 at.% Sm, this system can form metallic glasses or nanocomposite materials with low-density-high-strength properties [13]. The evaluation of thermodynamic stability of relevant phases is necessary to understand the complicated phase selection of this system especially under supercooling, which is key to achieving the desired compositions and microstructures.

## 2 Computational details

All simulations are performed using the MD technique with a timestep of 2 fs, as implemented in LAMMPS GPU-accelerated package [14,15]. Systems are fully equilibrated in 500,000 timesteps in canonical ensemble ($NVT$) or isothermal-isobaric ensemble ($NPT$) with the Nose-Hoover thermostat [16,17]. The main purpose of performing MD simulations in this work is to calculate the ensemble average of certain quantities (details are shown below), which is equivalent to the temporal average under the ergodic hypothesis. The average is collected in another 500,000 timesteps after the equilibrium is reached. For efficient energy and force calculations, we use semi-empirical interatomic potential in the Finnis-Sinclair form [18], which was developed to reproduce pure Al properties, energetics of Al-Sm intermetallic alloys and Al-Sm liquid structures [19]. This potential was particularly designed to treat Al-rich alloys (at.% Sm < ~ 10%).

## 3 Pure fcc-Al and Al liquid

We start with the calculation of free energy of the fcc-Al phase with TI, using the Einstein crystal as a reference system. The Helmholtz free energy of a classical Einstein crystal can be determined analytically as $F_0 = 3Nk_BT \ln(h\nu/k_BT)$, with $N$ the number of atoms, $h$ the Planck constant, $\nu$ the vibrational frequency and $k_B$ the Boltzmann constant. To implement TI, one generates intermediate systems with potentials $U(\lambda) = (1-\lambda)U_E + \lambda U_{Al}$, where $U_E$ and $U_{Al}$ stand for the potentials for the Einstein crystal and the real Al system, respectively. Then, the difference in Helmholtz free energy between the two systems can be expressed as

$$F_{Al,s} - F_0 = \int_0^1 \langle \frac{dU(\lambda)}{d\lambda} \rangle_{\lambda,NVT} d\lambda = \int_0^1 \langle U_{Al} - U_E \rangle_{\lambda,NVT} d\lambda. \qquad (1)$$

In Eq. (1), the subscript $s$ stands for solid, and $\langle \cdots \rangle_{\lambda,NVT}$ denotes the canonical ensemble ($NVT$) average of fcc-Al with respect to the intermediate potential $U(\lambda)$. The volume is fixed at the equilibrium volume at ambient pressure, which is determined separately via MD simulation with the real FS potential for Al under $NPT$ conditions. In this way, the Helmholtz free energy is equal to the Gibbs free energy at the same temperature.

As an example, we show in Fig. 1 the integrand of Eq. (1) for the implementation of TI at 800 K. The vibrational frequency $\nu$ for the Einstein crystal is chosen to be 5 THz, which is close to the principal peak of Al phonon density of states [20]. The integration, performed based on cubic spline interpolation of discrete data points collected by separate MD runs (red open circles), gives the free energy difference between fcc-Al and Einstein crystal reference $\Delta F = -3.872$ eV/atom.

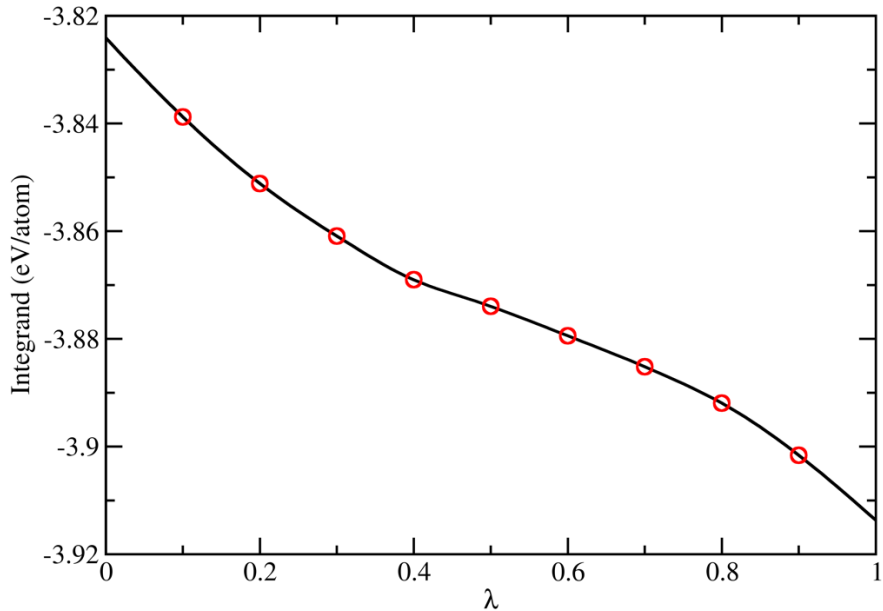

**Fig. 1.** The integrand of Eq. (1) for fcc Al at T = 800 K. Open circles are data points collected in separated MD runs. The solid line is a cubic spline interpolation.

To calculate the free energy of Al liquid, we first determine the melting point ($T_m$) of fcc-Al under ambient pressure, at which the difference in Gibbs free energy between the solid and liquid phases $\Delta G = 0$. Following the method described in Ref. [21], we plot the solid-liquid interface (SLI) velocity, obtained from MD simulation for the [100] direction, as function of temperature (see Fig. 2). The melting temperature determined from these data is 915.7±0.5 K, which is slightly lower than the experimental value (933 K). The Gibbs free energy difference at other temperatures is readily available by integrating the Gibbs-Helmholtz equation

$$\left[\frac{\partial(\Delta G/T)}{\partial T}\right]_P = -\frac{\Delta H}{T^2}, \quad (2)$$

where $\Delta H$ is the enthalpy change in the liquid and solid phases, or, the latent heat. The absolute free energy for Al liquid can be obtained by combining the information on solid-liquid free energy difference and the absolute free energy for the solid fcc-Al calculated previously. The final results are shown in Fig. 3.

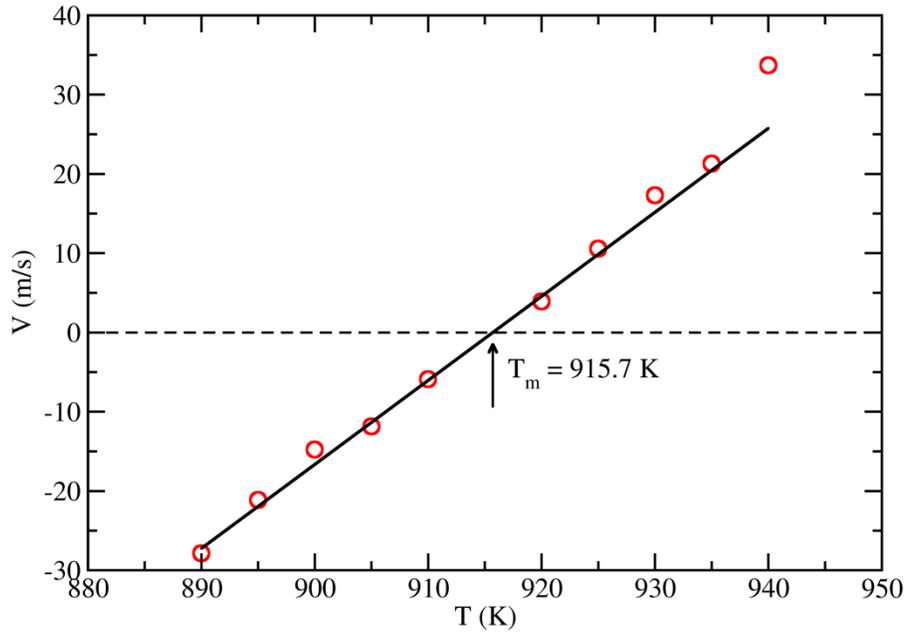

**Fig. 2.** The solid-liquid interface velocity for pure Al in the [100] direction as a function of temperature.

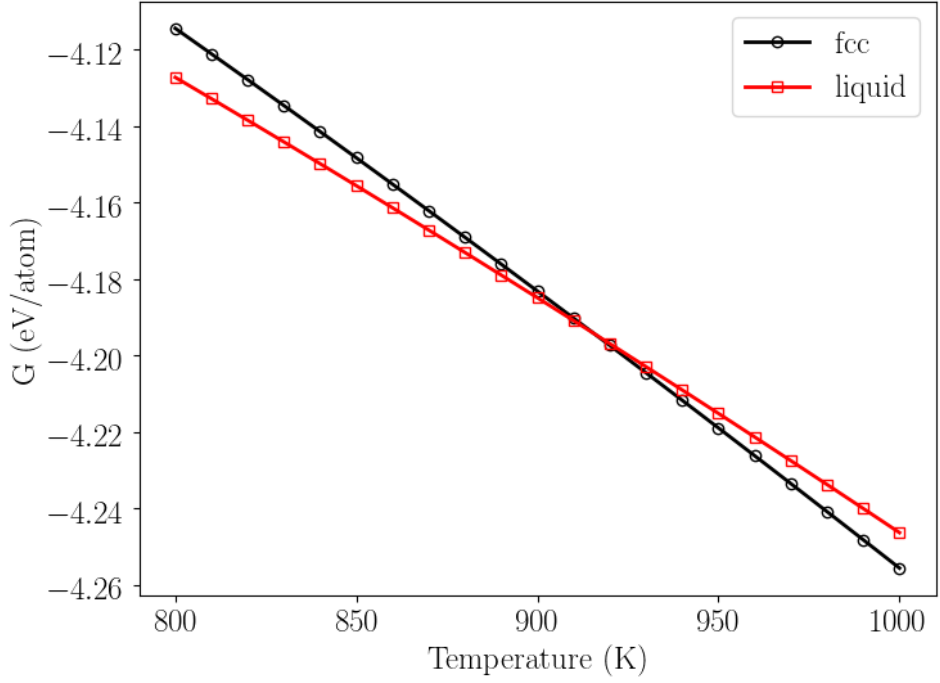

**Fig. 3.** Absolute Gibbs free energy of fcc and liquid Al as a function of temperature under the ambient pressure.

## 4 $Al_{1-x}Sm_x$ liquid

First, we introduce an auxiliary system $Al_{1-x}Sm'_x$, in which the factitious Sm′ atom has the mass of Sm but interacts with other atoms as Al. Thus, the Hamiltonian of $Al_{1-x}Sm'_x$ can be written as

$$H(p,q) = \sum_{i=1}^{3N(1-x)} \frac{p_i^2}{2m_{Al}} + \sum_{i=1}^{3Nx} \frac{p_i^2}{2m_{Sm}} + U_{Al}(q), \quad (3)$$

where $N$ is the number of atoms in the system, $m_{Al}$ and $m_{Sm}$ are the mass for Al and Sm atoms, respectively, and $(p,q)$ refers to a point in the phase space $(p_1, p_2, \cdots, p_{3N}, q_1, q_2, \cdots, q_{3N})$. The Helmholtz free energy for the liquid phase is defined as

$$F_{AlSm'} = -k_B T \ln\left\{\frac{1}{[N(1-x)]!\,(Nx)!\,h^{3N}} \int dp dq\, e^{-\beta H(p,q)}\right\}$$

$$= -k_B T \ln\left\{\frac{1}{[N(1-x)]!\,(Nx)!\,\Lambda_{Al}^{3N(1-x)}\Lambda_{Sm}^{3Nx}} \int dq\, e^{-\beta U_{Al}(q)}\right\}, \quad (4)$$

where $\beta = 1/k_B T$, and $\Lambda_\alpha$ is the de Broglie wavelength for species $\alpha$, which is defined as $\Lambda_\alpha = (h^2/2\pi m_\alpha k_B T)^{1/2}$. For pure Al liquid,

$$F_{Al} = -k_B T \ln\left[\frac{1}{N!\Lambda_{Al}^{3N}} \int dq e^{-\beta U_{Al}(q)}\right]. \quad (5)$$

Comparing Eqn. (4) and (5), one can obtain the Helmholtz free energy difference between $Al_{1-x}Sm'_x$ liquid and pure Al liquid:

$$F_{AlSm'} - F_{Al} = Nk_B T\left[\frac{3}{2}x \ln\frac{m_{Al}}{m_{Sm}} + x \ln x + (1-x)\ln(1-x)\right]. \quad (6)$$

Since $Al_{1-x}Sm'_x$ liquid and pure Al liquid share the same interaction potential, the equilibrium volume should also be the same under the same pressure. Thus, Eq. (6) also describes the Gibbs free energy difference between the two systems (the $PV$ term cancels out).

Next, we use TI to transform the factitious $Al_{1-x}Sm'_x$ system to the real $Al_{1-x}Sm_x$ system. To do that, we introduce intermediate systems interacting as $U(\lambda) = (1-\lambda)U_{Al} + \lambda U_{AlSm}$. Then, the difference in Gibbs free energy between the two systems can be expressed as

$$G_{AlSm} - G_{AlSm'} = \int_0^1 \langle\frac{dU(\lambda)}{d\lambda}\rangle_{\lambda,NPT} d\lambda = \int_0^1 \langle U_{AlSm} - U_{Al}\rangle_{\lambda,NPT} d\lambda, \quad (7)$$

where $\langle\cdots\rangle_{\lambda,NPT}$ stands for the isothermal-isobaric ($NPT$) ensemble average with respect to the intermediate potential $U(\lambda)$.

We use $x = 0.25$ and $T = 1500$ K as an example to describe the free energy calculation of liquid $Al_{1-x}Sm_x$ alloys. The transformation from Al liquid into the factitious Al-Sm' liquid results in a free energy change $\Delta F = -0.156$ eV/atom, as calculated according to Eq. (6). The implementation of TI to transform Al-Sm' into the real Al-Sm system is shown in Fig. 4, which gives $G_{AlSm} - G_{AlSm'} = -0.404$ eV/atom. Thus, the net difference of Gibbs free energy between the $Al_{0.75}Sm_{0.25}$ liquid and pure Al liquid is $-0.560$ eV/atom.

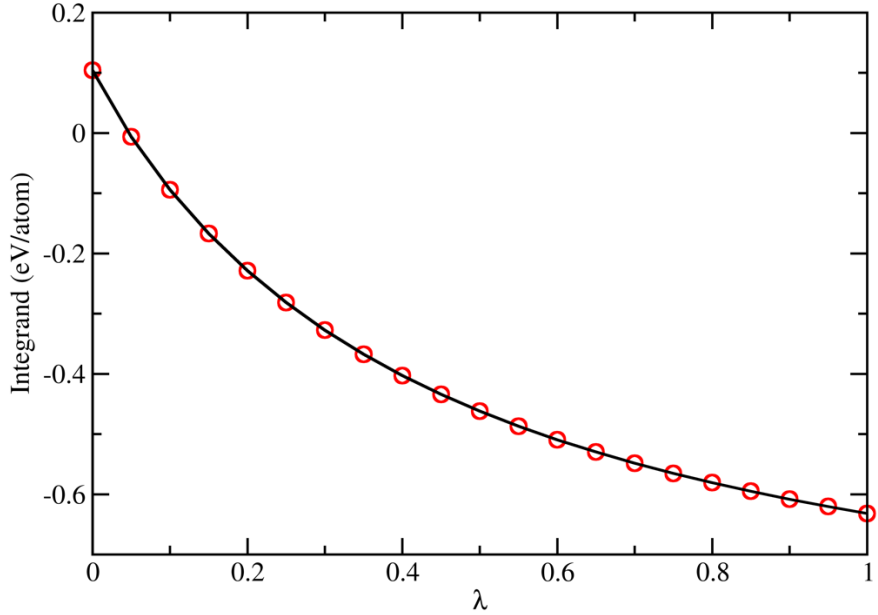

**Fig. 4.** The integrand of Eq. (7) for $x = 0.25$ and $T = 1500$ K. Open circles are data points collected in separated MD runs. The solid line is a cubic spline interpolation.

## 5 Applications

In the following, we demonstrate two applications of free energy calculations outlined in the above, namely, the determinations of driving force for fcc-Al nucleation in supercooled $Al_{1-x}Sm_x$ liquid and the melting curve for fcc-Al and $Al_3Sm$ crystals.

### 5.1 Driving force for nucleation of fcc-Al in supercooled $Al_{1-x}Sm_x$ liquid

Crystal nucleation in supercooled liquid is an important process in numerous areas of physical science [22]. It is also an important factor for glass formation, since glass is formed by suppressing crystal nucleation during fast quenching. As a marginal glass former, the glass formability of Al-Sm has a strong dependence on the Sm concentration [23]. When as-quenched Al-Sm glass is gradually heated, the devitrification process often starts with the deposit of Al nanocrystals [24,25]. Thus, study of the effect of Sm concentration on Al nucleation in supercooled Al-Sm liquids can provide useful information for both glass formation and devitrification processes [26]. The driving force is a fundamental parameter that describes the net bulk free energy gain upon the formation of a crystalline nucleus. For fcc-Al nucleation in supercooled $Al_{1-x}Sm_x$ liquid, the driving force can be expressed as

$$\Delta\mu = x\frac{\partial G_l}{\partial x} + G_s - G_l, \qquad (8)$$

where $G_l$ and $G_s$ refer to the Gibbs free energy of $Al_{1-x}Sm_x$ liquid and fcc-Al, respectively. In Fig. 5, we plot $\Delta\mu$ as a function of $x$ at a temperature of 700 K, where one can see that the driving force decreases as the Sm composition increases, but remains negative within the range of $x < 0.12$, showing that nucleation of fcc-Al is thermodynamically favored within this composition range. However, it should be noted that fcc-Al is the only solid phase considered in Fig. 5. When $x$ becomes large (before reaching 0.12), nucleation of other solid phases such as $Al_3Sm$ will become thermodynamically more favorable than fcc-Al.

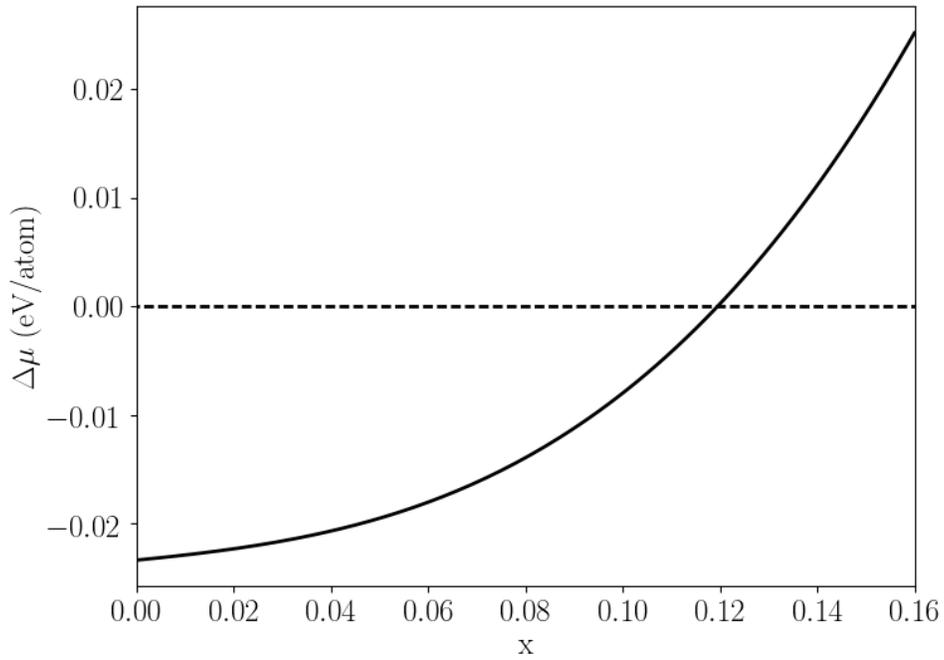

**Fig. 5.** Driving force for nucleation of fcc-Al in $Al_{1-x}Sm_x$ liquid as a function of the Sm composition.

### 5.2  Melting curve (liquidus) for fcc-Al and $Al_3Sm$

We also perform the free-energy calculation for the hexagonal $Al_3Sm$ phase, and traced out the melting curve (liquidus) for both fcc-Al and $Al_3Sm$. We focus on the technologically important Al-rich region for this system, in which fcc-Al and $Al_3Sm$ are the only two relevant solid phases according to the Al-Sm phase diagram [27,28]. Each coexistence point on the melting curve $(x, T)$ of a solid phase denotes a coexistence state, which satisfies

$$(x - x_s)\frac{\partial G_l(x,T)}{\partial x} + G_s(T) = G_l(x,T),$$

where $G_l$ and $G_s$ are the Gibbs free energy of the liquid and solid phases, respectively, and $x_s$ is the Sm composition in the solid phase. Mathematically, the coexistence composition at a specific temperature can be determined by the "tangent" construction as shown in Fig. 6, in which the formation Gibbs free energy of $Al_{1-x}Sm_x$ liquid $G_f$ is plotted as a function of the Sm composition $x$, at a supercooled temperature of 880 K. $G_f$ is calculated using the Gibbs free energy of fcc-Al and $Al_3Sm$ at the same temperature as reference states. In this way, $G_f$ for the two solid phases is zero (see Fig. 6). We construct tangential lines from the fcc-Al and $Al_3Sm$ phases to the liquid curve, shown as the red and blue lines in Fig. 6, respectively. From the position of the tangential points, one can determine the coexistence liquid composition with the two solid phases to be 0.039 and 0.054, respectively.

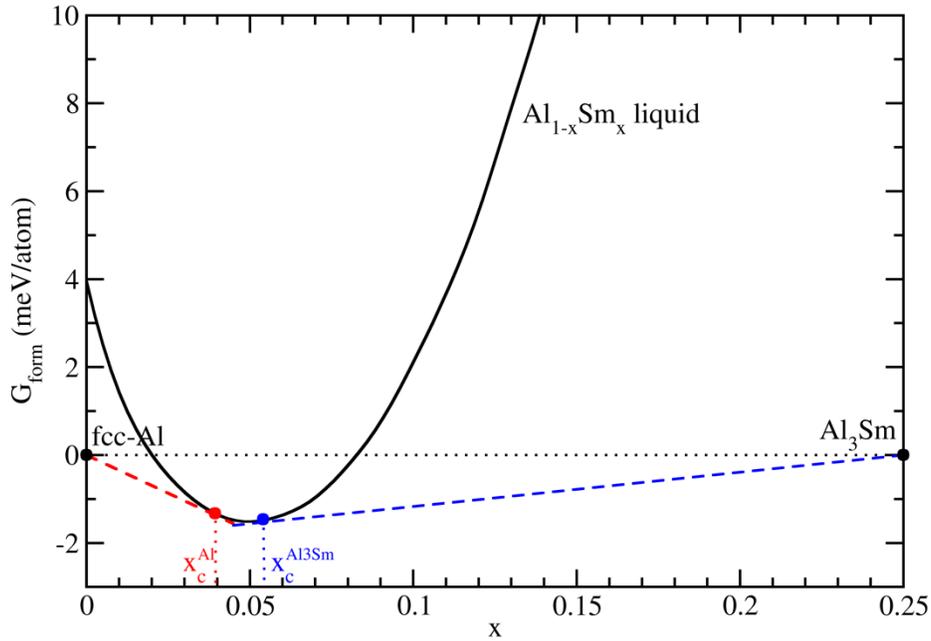

**Fig. 6.** The formation free energy of $Al_{1-x}Sm_x$ liquid, referenced to solid phases fcc-Al and $Al_3Sm$, at $T = 880$ K. The red and blue lines denote the tangential lines from fcc-Al and $Al_3Sm$ to the liquid curve, respectively. The tangential points give the coexistence liquid composition with the two solid phases, respectively.

The above procedure is repeated for various other temperatures to map out the melting curve for fcc-Al and $Al_3Sm$, as shown in Fig. 7. Our calculations predict a eutectic point at $T = 863$ K and $x = 0.051$, while the eutectic point from previous experiments [27,28] is located at $T = 908$

K and $x = 0.03$. At compositions away for the calculated eutectic point, our calculations generally underestimate the liquidus temperature by less than 100 K. Since the only energetics data used in fitting the Al-Sm FS potential was generated by density-functional theory (DFT) calculations at 0 K [19], we do expect some discrepancy with experiments in thermodynamic properties at finite temperatures. In this regard, a systematic way of determining solid-liquid phase equilibria, as outlined in the current paper, is valuable if one wants to refine a classical potential in order to more faithfully reproduce experimental thermodynamic information.

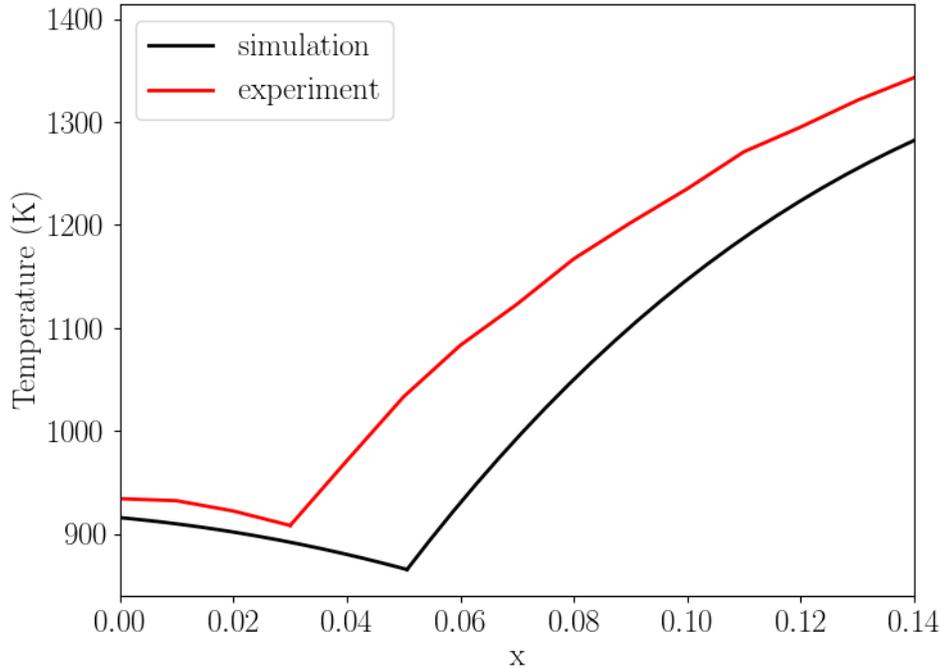

**Fig. 7.** Melting curve for fcc-Al and Al$_3$Sm from experiments [27,28] and the current calculations for $Al_{1-x}Sm_x$ in the Al-rich region of the Al-Sm system.

## 6  Conclusions

We establish a self-contained algorithm to rigorously evaluate the free energy for solid and liquid phases of an alloy system, based on thermodynamic integration. The algorithm starts from calculating the free energy of a single-element solid phase by referencing to a harmonic crystal. By monitoring the solid-liquid interface migration at different temperatures, we determine the melting point of the solid phase, which establishes a state of equality between the solid and liquid free energy. The free energy difference at other state points between solid and liquid phases can be obtained by integrating $\Delta H/T^2$ with temperature, where $\Delta H$ is the latent heat during melting.

Then, we generate an alchemical path connecting a pure liquid to a liquid alloy, and use thermodynamic integration to evaluate the mixing enthalpy and entropy. As an example, we apply this method on the Al-Sm system to determine the driving force for Al nucleation in Al-Sm liquid and the melting curve for the solid phases Al and $Al_3Sm$.

**Acknowledgements**

This work was supported by the Laboratory Directed Research and Development (LDRD) Program of The Ames Laboratory under the U.S. Department of Energy Contract No. DE-AC02-07CH11358.